\documentclass[11pt,aps,prx,onecolumn,superscriptaddress,amsmath,amssymb,floatfix,tightenlines]{revtex4-2}

\usepackage{CJK}
\usepackage{newtxtext}
\usepackage{graphicx}
\usepackage{bm}
\usepackage{siunitx}
\usepackage[sort&compress]{natbib}
\usepackage{xspace}
\usepackage{derivative}
\usepackage[nopartial]{fixdif}
\usepackage{physics2}
\usephysicsmodule{ab, ab.legacy}
\definecolor{linkblue}{RGB}{40,53,142}
\definecolor{tab_blue}{HTML}{1F77B4}
\usepackage{hyperref}
\hypersetup{
    colorlinks=true,
    linkcolor=tab_blue,
    urlcolor=tab_blue,
    citecolor=tab_blue,
    pdfpagemode=UseNone,
    linkbordercolor=0 1 0,
    pdfstartview=
}

\usepackage[]{cleveref}
\crefname{figure}{Fig.}{Figs.}
\Crefname{figure}{Figure}{Figures}
\crefname{table}{table}{tables}
\Crefname{table}{Table}{Tables}
\crefname{equation}{Eq.}{Eqs.}
\Crefname{equation}{Equation}{Equations}
\crefname{section}{Sec.}{Secs.}
\Crefname{section}{Section}{Sections}
\crefname{subsection}{Sec.}{Sec.}
\crefrangeformat{equation}{(#3#1#4)--(#5#2#6)}

\newcommand{\grad}{\bm{\nabla}}
\newcommand{\divergence}{\bm{\nabla}\cdot}
\newcommand{\iu}{\mathrm{i}}
\newcommand{\kBT}{k_\mathrm{B}T}
\newcommand{\Rey}{\ensuremath{\mathrm{Re}}\xspace}
\newcommand{\Reyeff}{\ensuremath{\mathrm{Re}_{\text{eff}}}\xspace}
\newcommand{\ueq}{u_{\text{eq}}}
\newcommand{\bx}{\bm{x}}

\newcommand{\bk}{\bm{k}}
\newcommand{\bp}{\bm{p}}
\newcommand{\bq}{\bm{q}}
\newcommand{\uhat}{\hat{u}}
\newcommand{\bu}{\bm{u}}
\newcommand{\buhat}{\hat{\bu}}
\newcommand{\bv}{\bm{v}}
\newcommand{\bN}{\bm{N}}
\newcommand{\bP}{\bm{\mathcal{P}}}
\newcommand{\cW}{\mathcal{W}}
\newcommand{\bcW}{\bm{\mathcal{W}}}
\newcommand{\bcWhat}{\widehat{\bm{\mathcal{W}}}}
\newcommand{\cZ}{\mathcal{Z}}
\newcommand{\bcZ}{\bm{\mathcal{Z}}}
\newcommand{\bcZhat}{\widehat{\bm{\mathcal{Z}}}}

\newcommand{\nueff}{\nu_{\text{eff}}}

\newcommand{\tv}{t_{\nu}}
\newcommand{\corr}{C}
\newcommand{\corrhat}{\widehat{C}}
\newcommand{\autocorr}{C_{v}}

\usepackage{soul}

\begin{document}

\title{\textbf{
Revisiting the scale dependence of the Reynolds number in correlated fluctuating fluids 
}}

\begin{CJK*}{UTF8}{gbsn}
\author{Sijie Huang (黄斯杰)}
\author{Ayush Saurabh}
\affiliation{
Department of Physics, Arizona State University, Tempe, AZ 85287, USA
}
\affiliation{
Center for Biological Physics, Arizona State University, Tempe, AZ 85287, USA
}
\author{Steve Press\'e}
\email{Corresponding author: spresse@asu.edu}
\affiliation{
Department of Physics, Arizona State University, Tempe, AZ 85287, USA
}
\affiliation{
Center for Biological Physics, Arizona State University, Tempe, AZ 85287, USA
}
\affiliation{
School of Molecular Sciences, Arizona State University, Tempe, AZ 85287, USA
}

\date{\today}

\begin{abstract}
For the incompressible Navier--Stokes equation, the Reynolds number ($\Rey$) is a dimensionless parameter quantifying the relative importance of inertial over viscous forces. In the low-$\Rey$ regime ($\Rey \ll 1$), the flow dynamics are commonly approximated by the linear Stokes equation. Here we show that, within the framework of spatially fluctuating hydrodynamics, this linearization breaks down when the thermal noise is spatially correlated, even if $\Rey \ll 1$. We perform direct numerical simulations of spatially correlated fluctuating hydrodynamics in both one and two dimensions. In one dimension, the linearized dynamics exhibit significantly slower relaxation of high-wavenumber Fourier modes than the full nonlinear dynamics. In two dimensions, an analogous discrepancy arises in the particle velocity autocorrelation function, which decays more slowly in the correlated linear Stokes case than in the correlated nonlinear Navier--Stokes case. In both settings, spatial correlations inhibit viscous momentum diffusion at small scales, leading to prolonged relaxation under the linear dynamics, whereas nonlinear mode coupling accelerates small-scale relaxation. Thus, the interplay between nonlinear coupling and viscous damping becomes scale dependent, invalidating the use of a single global Reynolds number. Taken together, these findings show that, for spatially correlated fluctuating fluids, the effective Reynolds number must be reinterpreted as a scale-dependent quantity.
\end{abstract}

\maketitle

\end{CJK*}

\section{Introduction\label{sec:1}}

The incompressible Navier--Stokes equation provides a fundamental description of Newtonian fluids at low speeds where density variations are negligible, encapsulating the interplay between inertial and viscous forces. The relative importance of these forces is quantified by the dimensionless control parameter known as the Reynolds number (\(\Rey\))~\cite{tamburrino2025,saldana2024fluids}. In regimes where $\Rey\ll1$---common in systems such as bacterial locomotion, sedimentation, and microfluidic transport---flows are slow or highly viscous, and inertia plays no significant role~\cite{berg1996,groisman2004,waigh2024cup,davis1985arfm}. Accordingly, in this regime, the nonlinear convective term in the Navier--Stokes equation can be omitted, reducing the dynamics to the linear Stokes equation~\cite{batchelor2000,tamburrino2025}. This linearization has been extensively validated, and the resulting Stokes approximation underpins a broad range of analyses and applications across a wide range of fields, such as biophysics, soft matter, and geophysics~\cite{qiu2014nc,baskaran2009,salac2012jfm,he2024prf,nguyen2005jfm,rallabandi2024arfm,agarwal2025pof,brinkerhoff2021,pattyn2003}. A notable example of the success of the Stokes equation is its application to the classical Brownian motion~\cite{kim1991,hinch1975}. As the Stokes equation is linear, the flow field generated by a point force can be expressed analytically through its Green's function, the Stokeslet, which forms the foundation of Stokesian dynamics~\cite{brady1988,phung1996,banchio2003}. This framework provides a hydrodynamically consistent description of colloidal and Brownian motion by resolving fluid-mediated interactions and enforcing momentum conservation, thereby capturing long-time correlations and many-body effects absent in local Langevin descriptions and not automatically reproduced by phenomenological generalized Langevin models~\citep{michael2025pof,atzberger2007,ouaknin2021,fiore2019,seyler2020,seyler2019}.

Despite the success of the linear Stokes approximation, the conventional low-\Rey argument presumes that the standard Laplacian form of viscous diffusion remains appropriate at all relevant scales, under which viscous damping strengthens at small scales and dominates inertial effects. This, in turn, justifies the use of a single, global \Rey as a criterion for the validity of the linear approximation. When the assumption of standard Laplacian is relaxed---for instance, in fluids with spatial structures---the resulting modification of viscous damping across scales alters the relative importance of inertia and viscosity, thereby challenging the conventional low-\Rey argument for the validity of the linear approximation. 

In standard fluctuating hydrodynamics, the low-\Rey argument remains well justified, in which Stokesian dynamics emerges as the linear, low-\Rey limit of a more general theory. 
Here, fluctuating hydrodynamics, originally formulated by \citet{landau1959}, provides a coarse-grained description of fluids at mesoscopic scales---large compared to the molecular mean free path yet small enough that thermal fluctuations remain significant~\cite{kavokine2021arfm}. In this formalism, thermal fluctuations are introduced into the incompressible Navier--Stokes equation as a random stress satisfying a fluctuation--dissipation relation (FDR), which ensures that viscous dissipation and thermal energy injection are balanced at every spatial scale, thereby maintaining thermal equilibrium~\cite{zarate2006}. The linear Stokes approximation within fluctuating hydrodynamics remains well justified, since the viscous diffusion retains its standard Laplacian form, and numerous theoretical analyses and numerical simulations have demonstrated that this linearization accurately captures various diffusive transport phenomena, even under nonequilibrium conditions~\cite{delmotte2025,turk2024jfm,hauge1973jsp,donev2014jsm,usabiaga2013jchemphys,delmotte2015,noetinger1990}.

In the standard Landau--Lifshitz formulation, thermal noise is assumed to be white in both space and time. However, in complex fluids---such as viscoelastic or otherwise structured fluids---microscopic stresses can become spatially correlated due to microstructural coupling or slow stress relaxation~\cite{grimm2025sm}. 
Motivated by these observations, we have recently generalized the Landau--Lifshitz formulation by allowing the thermal noise to exhibit spatial correlations~\cite{huang2025}. In this generalized setting, the FDR requires matching correlations in the viscous term, thereby modifying the standard Laplacian form of viscous dissipation, rendering it nonlocal and leading to a scale-dependent effective viscosity. Consequently, viscous diffusion---and therefore the relative balance between inertial and viscous effects---varies across scales. The extent of this variation is determined by the form of the spatial correlation, suggesting that the conventional, single-valued Reynolds number may no longer suffice. Instead, spatial correlations introduce a continuous spectrum of scale-dependent Reynolds numbers.

These considerations naturally raise a fundamental question: how do spatial correlations, when consistently incorporated through the FDR, alter the competition between inertia and viscous effects across scales? In particular, at thermal equilibrium, does the classical low-\Rey justification for the linear Stokes approximation remain valid when the thermal noise is spatially correlated? Addressing this question requires revisiting the physical meaning of the Reynolds number itself when viscosity and momentum transport become scale dependent, thereby motivating this work.

To address this question, we analyze how spatial correlations modify the relative strength of inertia and viscous dissipation across spatial scales. As spatial correlations in the viscous term modify how momentum spreads through the fluid, velocity fluctuations at different wavelengths relax at different rates. We therefore examine the dynamics in Fourier space. As such, we show that spatial correlations give rise to a scale-dependent effective Reynolds number, \(\Reyeff(k)\), which varies continuously with wavenumber. This effective Reynolds number is dictated by the spatial correlation function and makes explicit how the balance between inertia and viscosity depends on scale. In uncorrelated fluids, \(\Reyeff\) reduces to the conventional, scale-independent Reynolds number. In correlated fluids, however, small-scale viscous damping weakens as correlations suppress momentum diffusion, making inertial effects relatively stronger even when the conventional Reynolds number is small. As a result, the dynamics can no longer be described by the linear Stokes approximation, demonstrating that the classical notion of a single scale-independent Reynolds number is insufficient for spatially correlated fluctuating fluids.

Having established this theoretical framework, we validate it through direct numerical simulations in both one and two dimensions. The one-dimensional system is computationally inexpensive, allowing us to collect sufficient statistics to examine the relaxation of individual Fourier modes directly, thereby providing a clean test of the predicted suppression of small-scale viscous damping under spatial correlations. In two dimensions, where collecting sufficient Fourier-mode statistics is computationally prohibitive, we instead analyze tracer-particle diffusion, whose velocity autocorrelation function (VACF) reflects the cumulative influence of momentum relaxation across all fluid modes. In both settings, we compare the linearized dynamics with the full nonlinear dynamics for varying correlation lengths. Consistent with the theoretical predictions, we find that spatial correlations slow relaxation under linear dynamics and that nonlinear mode coupling accelerates small-scale relaxation. Taken together, these results demonstrate that spatial correlations render inertia--viscosity competition scale-dependent, invalidating the conventional use of a single, scale-independent Reynolds number.

\begin{figure}
    \centering
    \includegraphics[width=\linewidth]{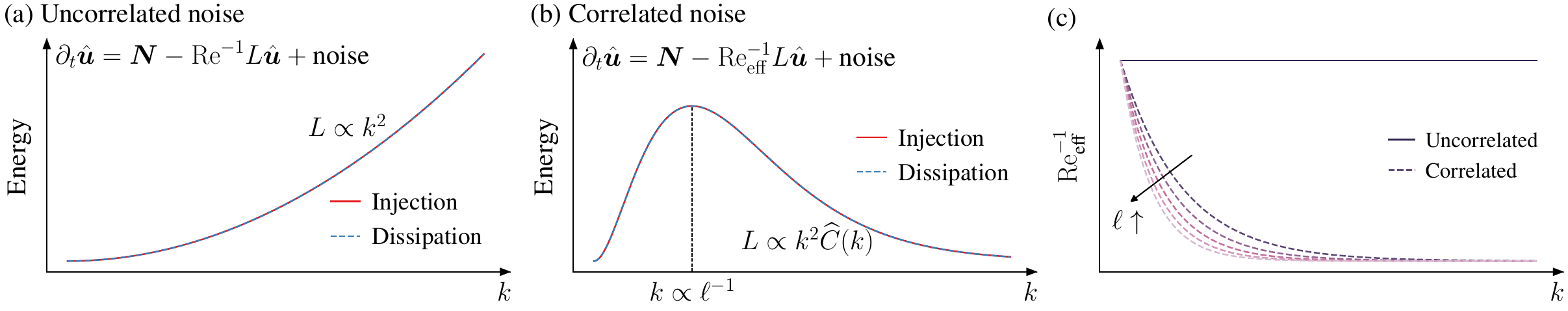}
    \caption{Scale dependence of the Reynolds number in the spatially correlated fluctuating Navier--Stokes equation. (a) For spatially uncorrelated noise, both the noise energy injection and viscous dissipation spectra scale as $k^2$. Therefore, they balance exactly, consistent with the FDR. (b) For spatially correlated noise, the correlation function $\corrhat(k)$ modifies both the spectra of energy injection and viscous dissipation, preserving their balance and the FDR. (c) The prefactor of the viscous term remains constant for spatially uncorrelated noise but decreases with increasing \(k\) for spatially correlated noise, indicating that the viscous term becomes less important at large wavenumbers. Here, $\buhat$ is the Fourier mode at wavevector $\bk$, $\bm{N}$ is the nonlinear term, $L$ is the dissipative linear operator, $\ell$ is the characteristic length scale of the correlation function $\corr(r)$.
    }
    \label{fig:1}
\end{figure}

The rest of this paper is organized as follows. \Cref{sec:2} briefly reviews the spatially correlated fluctuating hydrodynamics introduced in Ref.~\cite{huang2025} and summarizes the modifications required by the FDR. \Cref{sec:3} analyzes the resulting nonlocal momentum diffusion in Fourier space and introduces a scale-dependent Reynolds number that quantifies how spatial correlations reshape the balance between inertial and viscous effects. \Cref{sec:4} presents numerical simulations in one and two dimensions that test these theoretical predictions by comparing linearized and fully nonlinear dynamics across correlation lengths. \Cref{sec:5} summarizes our findings and discusses their broader implications for fluctuating hydrodynamics and correlated complex fluids.

\section{Spatially Correlated Fluctuating hydrodynamics\label{sec:2}}

In this section, we briefly review the spatially correlated fluctuating hydrodynamics introduced in Ref.~\cite{huang2025}. We begin with the standard Landau--Lifshitz formulation and allow the thermal noise to exhibit spatial correlations. Enforcing the FDR to maintain thermal equilibrium then yields a spatially correlated fluctuating incompressible Navier--Stokes equation.

At the continuum level, thermally fluctuating fluids can be described by the fluctuating incompressible Navier--Stokes equation,
\begin{equation}
    \label{eq:llns}
    \partial_t\bu + \bu\cdot\grad\bu = -\grad p + \nu\nabla^2\bu + \sqrt{2\nu\Theta}\,\divergence\bcW,
\end{equation}
where $\bu$ is the fluid velocity, $p$ the pressure, \(\nu\) the kinematic viscosity, and the noise amplitude \(\Theta=\kBT/\rho\) is proportional to the thermal energy at temperature \(T\) and density \(\rho\). The velocity field \(\bu\) is interpreted as a coarse-grained description of fluid motion on length scales larger than the molecular mean free path~\cite{bandak2022,bandak2024}. For incompressible, isothermal fluids, $\rho$, $\nu$, and $T$ are taken as constants. The random stress \(\bcW\) models the thermal noise as a spatiotemporal white Gaussian field with covariance~\cite{zarate2006}
\begin{equation}
    \label{eq:uncorrelated_noise}
    \aab*{\cW_{ij}(\bx,t)\cW_{lm}(\bx',t')} = (\delta_{il}\delta_{jm} + \delta_{im}\delta_{jl})\delta(r)\delta(t-t'),
\end{equation}
where $r=|\bx-\bx'|$ is the separation. The fluctuations in the velocity field \(\bu\) are driven by the divergence of this random stress, \(\divergence\bcW\). When expressed in Fourier space, this stochastic forcing appears as \(\bk\cdot\bcWhat(\bk)\), whose variance scales as \(k^2\). Consequently, the noise injects energy into each velocity mode at a rate proportional to \(k^2\). The viscous Laplacian \(\nabla^2\) also dissipates energy at a rate proportional to \(k^2\), so that energy injection and dissipation balance exactly at every wavenumber. As the nonlinear convective term conserves energy, it does not contribute to this scale-by-scale energy injection--dissipation balance~\cite{bandak2022}. As a result, the FDR is determined entirely by the matching of noise injection and viscous dissipation for each wavenumber, as illustrated in~\cref{fig:1}(a), which ensures thermal equilibrium. Indeed, it has been shown that this classical fluctuating hydrodynamic formulation yields the correct equilibrium spectrum of velocity fluctuations, consistent with the Boltzmann distribution of the kinetic energy~\cite{bandak2022,usabiaga2012mms,donev2010camcs}.

In Ref.~\cite{huang2025}, we generalized \cref{eq:llns} to incorporate spatially correlated, temporally white noise $\bcZ(\bx,t)$ with covariance 
\begin{equation}
    \label{eq:correlated_noise}
    \aab{\cZ_{ij}(\bx,t)\cZ_{lm}(\bx',t')} = (\delta_{il}\delta_{jm} + \delta_{im}\delta_{jl})\corr(r)\delta(t-t'),
\end{equation}
where the correlation function $\corr(r)$ only depends on the separation \(r\) and characterized by a single length scale $\ell$. In this work, we consider a \(d\)-dimensional Lorentzian-type correlation function 
\begin{equation}
    \label{eq:lorentzian}
    \corr(r) \propto \frac{\ell}{(r^2 + \ell^2)^{(d+1)/2}},
\end{equation}
for physical relevance and analytical tractability. This correlation function approaches the \(d\)-dimensional delta function in the limit of \(\ell=0\), thus allowing a systematic deviation from the white-noise limit.

Spatial correlations in the noise redistribute energy injection across scales, so energy is no longer injected at a rate proportional to \(k^2\) in Fourier space. This disrupts the scale-by-scale energy balance required for thermal equilibrium. To restore this balance and maintain equilibrium, the viscous term must incorporate the same spatial correlation as the noise, so that the per-mode energy injection and viscous dissipation remain matched and the FDR is preserved, as shown in~\cref{fig:1}(b). This requirement follows jointly from the FDR and the condition that each velocity mode satisfies equipartition at equilibrium. A more detailed discussion of this requirement is provided in Ref.~\cite{huang2025}. In physical space, this leads to the spatially correlated fluctuating incompressible Navier--Stokes equation with nonlocal diffusion
\begin{equation}
    \label{eq:generalized_fns}
    \partial_t\bu + \bu\cdot\grad\bu = -\grad p + \divergence(\nueff*\grad\bu) + \sqrt{2\nu\Theta}\divergence\bcZ,
\end{equation}
where \(\nueff(r) = \nu C(r)\) is the effective viscosity, and \(*\) denotes convolution. The convolution indicates that momentum diffusion becomes nonlocal: the viscous response at a spatial location depends on the velocity gradients $\grad\bu$ in a surrounding region. This modification gives rise to a scale-dependent effective viscosity \(\nueff\) whose magnitude varies with the spatial structure of \(\corr(r)\). 

This nonlocal viscous term is not a modeling choice: its form is uniquely fixed by the combined requirements of thermodynamic consistency, the FDR, and energy equipartition. In combination with the correlated noise, it ensures that the equilibrium energy spectrum of velocity fluctuations remains consistent with the Boltzmann distribution. Any other form would violate these equilibrium constraints and drive the system out of equilibrium.

Consequently, viscous diffusion becomes scale dependent, and the competition between inertial and viscous effects can no longer be characterized by the conventional single-valued Reynolds number [\cref{fig:1}(c)]. Instead, spatial correlations determine how this balance varies with scale, motivating the introduction of a scale-dependent Reynolds number in the next section.

\section{Scale-dependent Reynolds number\label{sec:3}}

Spatial correlations in the thermal noise modify the viscous damping in \cref{eq:generalized_fns}, rendering momentum diffusion inherently nonlocal and scale dependent. To quantify how this modified viscous response affects the relative importance of inertia and viscosity, we analyze the dynamics in Fourier space and derive a scale-dependent Reynolds number that governs the behavior of each wavenumber mode. Unlike the conventional Reynolds number, which is a single global value set by macroscopic scales, this effective Reynolds number forms a continuous spectrum whose values are determined by the spatial correlation function.

\subsection{Fourier-space representation}

Taking the Fourier transform of \cref{eq:generalized_fns} yields
\begin{equation}
    \label{eq:generalized_fns_fourier}
    \odv*{\buhat(\bk)}{t} = \bP\bN(\bk) - \nu\corrhat(k)k^2\buhat(\bk) + \iu\sqrt{2\nu\Theta}\,\bP\bk\cdot\bcZhat(\bk),
\end{equation}
where the nonlinear convective term is 
\begin{equation}
    \bN(\bk) = -\iu\sum_{\bp+\bq=\bk}(\bk\cdot\buhat(\bq))\buhat(\bp),
\end{equation}
and the projection operator \(\bP=\mathbf{I} - \bk\bk^\top/k^2\), arising from \(\grad p\), ensures that the incompressibility condition \(\bk\cdot\buhat(\bk)=0\) is satisfied. The Fourier transform of the Lorentzian correlation in \cref{eq:lorentzian},
\begin{equation}
    \corrhat(k) = \exp(-\ell k),
\end{equation}
enters the viscous term directly and therefore modulates the damping rate. For uncorrelated noise (\(\ell=0\)), \(\corrhat(k)=1\), one recovers the standard Laplacian damping \(\nu k^2\). For correlated noise (\(\ell>0\)), \(\corrhat(k)\) suppresses viscous damping exponentially at large wavenumbers, resulting in significantly slower relaxation of short-wavelength modes.

At thermal equilibrium, the velocity fluctuations of a fluid are distributed according to the Boltzmann distribution, which assigns each independent incompressible Fourier mode the same average kinetic energy, a manifestation of energy equipartition~\cite{bandak2022}. Formally, this equipartition is expressed as~\cite{usabiaga2012mms}
\begin{equation}
    \label{eq:equipartition}
    \aab*{|\buhat(\bk)|^2} = (d-1)\Theta, 
\end{equation}
where \(d\) denotes the spatial dimension. The prefactor \(d-1\) reflects the removal of the longitudinal mode under incompressibility~\citep{usabiaga2012mms}. The equipartition provides a characteristic equilibrium velocity scale
\begin{equation}
    \ueq = \sqrt{(d-1)\Theta}.
\end{equation}
As this scale is independent of \(k\), all scale dependence in \cref{eq:generalized_fns_fourier} arises solely from the modified viscous damping through \(\corrhat(k)\). 

\subsection{Competition between inertial and viscous effects across scales}

To quantify the competition between nonlinear convection and viscous damping at each wavenumber, we compare their characteristic magnitudes in \cref{eq:generalized_fns_fourier} 
\begin{equation}
    \text{inertial}\sim k\ueq^2,\quad \text{viscous}\sim \nu k^2\corrhat(k)\ueq.
\end{equation}
Their ratio defines a scale-dependent effective Reynolds number,
\begin{equation}
    \label{eq:Re_eff}
    \Reyeff(k) = \frac{\ueq}{\nu k\corrhat(k)} = \frac{\ueq}{\nu k}e^{\ell k}.
\end{equation}
This definition generalizes the conventional single-valued Reynolds number to a scale-dependent quantity. For uncorrelated noise, \(\Reyeff(k)\) decays as \(k^{-1}\), and so the largest-scale wavenumber \(k_0\) uniquely controls the dynamics: if \(\Reyeff(k_0)\ll1\), then \(\Reyeff(k)\ll1\) for all \(k>k_0\), and the linear Stokes approximation is justified across scales. In this sense, the conventional Reynolds number can be identified as \(\Reyeff(k_0)\).

For correlated noise, the exponential factor \(e^{\ell k}\) dominates the algebraic \(k^{-1}\) decay at large wavenumbers, causing \(\Reyeff(k)\) to grow exponentially. Consequently, even when the conventional Reynolds number is small, sufficiently short-wavelength modes satisfy \(\Reyeff(k)\). Inertial effects, therefore, become increasingly important at these scales. The correlation length \(\ell\) controls how quickly this high-\(k\) growth sets in: larger \(\ell\) shifts the onset of the inertial regime to larger wavelengths and amplifies the scale-dependent departure from the conventional, single-valued Reynolds number.

A complementary and intuitive way to view this competition is via characteristic timescales. Define the inertial timescale and the viscous timescale for a mode at wavenumber \(k\) as 
\begin{equation}
    \tau_i = (k\ueq)^{-1},\quad \tau_v = (\nu k^2e^{-\ell k})^{-1}.
\end{equation}
Their ratio returns the effective Reynolds number, \(\Reyeff(k) = \tau_i^{-1}/\tau_v^{-1}\). For uncorrelated noise, \(\tau_v\sim k^{-2}\) decreases faster than \(\tau_i\sim k^{-1}\), so small-scale relaxation is dominated by viscous damping. For correlated noise, however, \(\tau_v\) grows exponentially, resulting in extremely slow decay of short-wavelength modes; increasing \(\ell\) accentuates this effect by further suppressing the viscous damping at high \(k\). Nonlinear mode coupling, which redistributes energy among the modes on the inertial timescale \(\tau_i\), then overwhelms the increasingly weak viscous damping at large \(k\). This mismatch between \(\tau_i\) and \(\tau_v\) is the physical mechanism underlying the breakdown of the linear approximation.

As viscous damping weakens so dramatically at high \(k\), the linearized Stokes equation no longer correctly captures the multiscale dynamics when \(\ell>0\). The conventional Reynolds number loses predictive power: even when it is small, inertia remains significant at sufficiently short wavelengths. Therefore, correlated fluctuating fluids cannot be characterized by the classical scale-independent Reynolds number. Instead, their behavior must be described by the full spectrum of scale-dependent Reynolds numbers \(\Reyeff(k)\), which reveal where nonlinear coupling is essential in restoring momentum transfer.

More broadly, according to \cref{eq:Re_eff}, whenever the correlation spectrum $\corrhat(k)$ decays faster than $k^{-1}$ at high wavenumbers, the effective viscous damping $\nu k^{2}\corrhat(k)$ falls off too rapidly to balance inertia, causing $\Reyeff(k)$ to grow. Thus, the failure of the linear Stokes approximation is not specific to the Lorentzian correlation considered here, but arises for any correlation function with sufficiently fast decay in $k$.

\section{Results\label{sec:4}}


To test the theoretical predictions developed in \cref{sec:3}, we perform direct numerical simulations of fluctuating hydrodynamics in both one and two spatial dimensions. The one-dimensional system, governed by the stochastic Burgers equation~\cite{delong2013pre}, provides a simplified and computationally tractable setting that allows us to directly measure the relaxation of each Fourier mode via its autocorrelation---an analysis that is prohibitively expensive in two dimensions. Although the stochastic Burgers equation lacks incompressibility and many other features of the Navier--Stokes dynamics, it preserves the essential competition between nonlinear advection and viscous dissipation, making it a one-dimensional analogue for studying fluctuating hydrodynamics~\cite{donev2010camcs,delong2013pre}. These one-dimensional results illustrate qualitatively how spatial correlations suppress viscous damping and lead to a breakdown of the linear approximation (\textit{i.e.}, the stochastic heat equation) at large wavenumbers. 

In two dimensions, we probe relaxation dynamics through tracer-particle diffusion. The particle VACF provides a sensitive and experimentally relevant measure of relaxation dynamics, serving as a practical surrogate for per-mode velocity statistics. 
These two-dimensional simulations represent physically realistic fluctuating flows, and allow us to test the breakdown of the linear Stokes approximation both qualitatively, through the VACF decay, and quantitatively, through deviations in the tracer diffusion coefficient. 

Across both dimensions, we systematically vary the correlation length $\ell$ that controls the Lorentzian correlation function in \cref{eq:lorentzian}, and compare the corresponding linearized dynamics with the full nonlinear dynamics to assess the conditions under which the conventional, single-valued Reynolds number fails. We begin with Fourier-mode autocorrelations in one dimension (\cref{sec:4.1}) and then analyze tracer dynamics in two dimensions (\cref{sec:4.2}).

We follow the simulation strategy as described in Ref.~\cite{huang2025} for both the one- and two-dimensional cases. We summarize the key aspects below; further details and simulation conditions are provided in the Supplemental Material~\cite{supplemental}. For spatial discretization, we employ a Fourier--Galerkin pseudospectral method with the $3/2$--rule de-aliasing to remove aliasing from the nonlinear terms~\cite{canuto2007}. For tracer diffusion in two dimensions, we follow the immersed-boundary method developed in Refs.~\cite{usabiaga2013jcp,usabiaga2014cmame}, with particle radius $a=\qty{0.05}{\um}$. For time integration, the nonlinear terms are advanced with Heun's method~\cite{usabiaga2012mms}, while the linear terms and stochastic forcings are treated exactly via integrating factors~\cite{canuto2006,oksendal2003}. Particle trajectories are updated using a midpoint predictor--corrector scheme~\cite{usabiaga2014cmame}. The one-dimensional simulations are carried out in a periodic domain of length $2\pi~\unit{\um}$, and the two-dimensional ones in a doubly periodic domain of side length $4\pi~\unit{\um}$. Periodic boundaries mimic an unbounded homogeneous fluid, eliminating boundary fluxes so that no net energy is injected or removed, thus ensuring the FDR holds without boundary corrections.

For the one-dimensional simulations, we take the viscosity to be $\nu = 875~\unit{\um^{2}/\ms}$, corresponding to water at \qty{300}{\K}. The conventional Reynolds number is defined as $\Rey = \ueq/(k_0\nu) \approx 0.2$, where $k_0=\qty{1}{\um}^{-1}$ is the wavenumber corresponding to the domain length. For the two-dimensional simulations, we additionally consider a lower viscosity value, $\nu = 100~\unit{\um^{2}/\ms}$, representative of liquid metals such as mercury around temperature \qty{293}{\K}. The corresponding conventional Reynolds numbers for these two viscosities are $\Rey\approx0.4$ and $3$, with $k_0=\qty{0.5}{\um}^{-1}$.

We summarize the main findings here before presenting the detailed analyses in the following subsections. In one dimension (\cref{sec:4.1}), we find that finite spatial correlations lead to a systematic separation between linearized and nonlinear dynamics: as shown in \cref{fig:2}, Fourier-mode autocorrelations decay more slowly in the linear regime than in the nonlinear one, with the discrepancy becoming more pronounced at large wavenumbers. In two dimensions (\cref{sec:4.2}), similar discrepancies between linearized and nonlinear dynamics are observed through tracer-particle diffusion. At $\Rey \approx 0.4$, \cref{fig:3} shows that particles exhibit stronger diffusion in the linear regime than in the nonlinear regime, manifested by a slower long-time decay of the VACF and a larger diffusion coefficient. At the largest correlation length considered, the diffusion coefficient predicted by the linear dynamics exceeds the nonlinear value by nearly $90\%$. Qualitatively similar behavior is observed at the higher Reynolds number $\Rey \approx 3$ in \cref{fig:4}. Taken together, these results reveal systematic differences between linearized and nonlinear fluctuating hydrodynamics that persist across dimensions and Reynolds numbers.

\subsection{Spatial correlations slow the relaxation of Fourier modes\label{sec:4.1}}

\begin{figure}
    \centering
    \includegraphics[width=\linewidth]{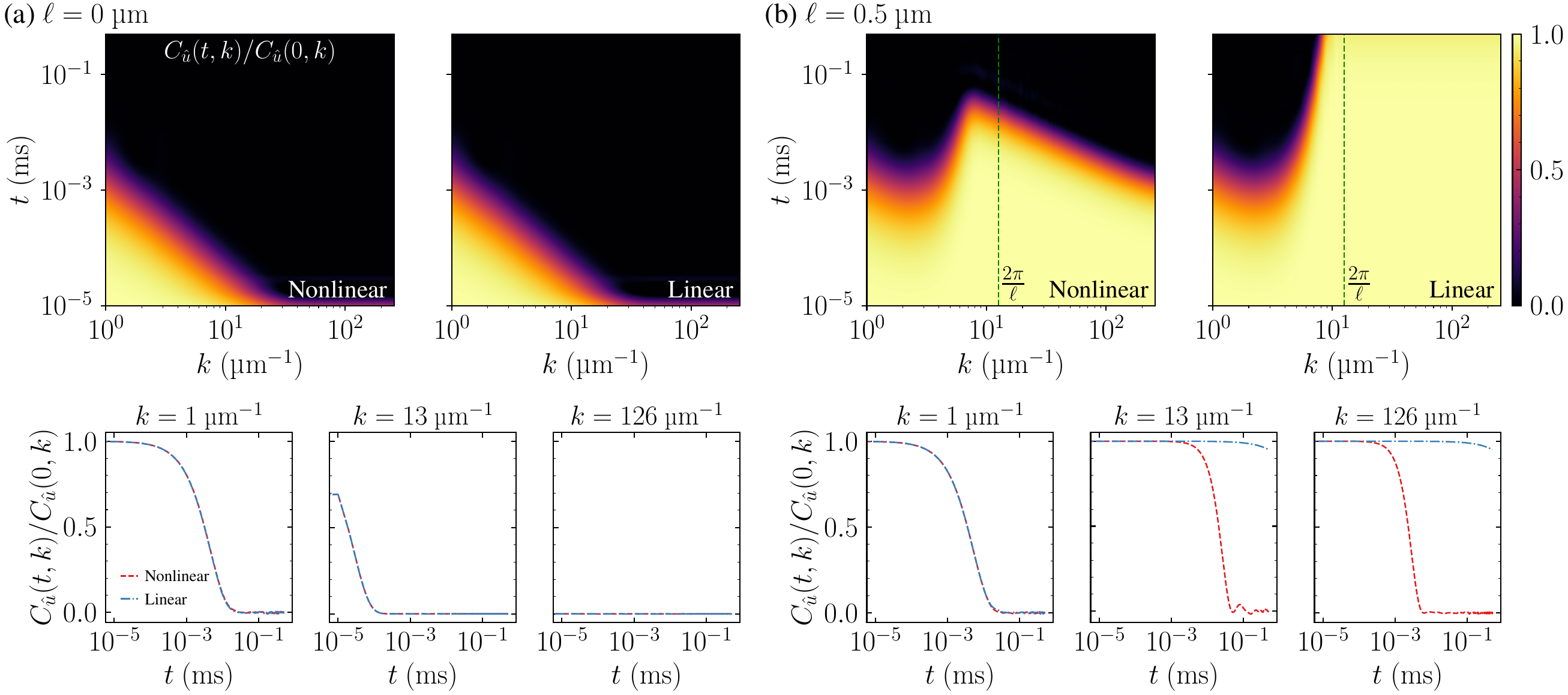}
    \caption{Nonlinear mode coupling in the Burgers equation accelerates velocity relaxation compared with the linear heat equation. (a) For white noise, the autocorrelation contours of the Fourier modes \(\uhat\) exhibit similar behavior in both the nonlinear (left) and linear (right) dynamics: low-wavenumber modes decay more slowly, whereas high-wavenumber modes relax more rapidly. The VACFs at three representative wavenumbers (\(k=1, 13,\) and \(126~\unit{\um}^{-1}\)) show great agreement between the nonlinear (Burgers) and linear (heat) cases. (b) For correlated noise, however, the nonlinear dynamics yield systematically faster relaxation, most notably at higher wavenumbers (\(k\gtrsim10~\unit{\um}^{-1}\)), where the linear dynamics exhibit near-frozen high-$k$ modes, illustrating the failure of the linear approximation. The VACFs at the same three wavenumbers in (a) confirm this observation.}
    \label{fig:2}
\end{figure}

We begin by examining the relaxation of individual Fourier modes in one dimension. We consider two correlation lengths, $\ell = 0$ and $0.5~\unit{\um}$. By comparing the linearized and nonlinear dynamics, we characterize how spatial correlations influence the decay of high-wavenumber modes and how nonlinear interactions modify this behavior. The relaxation is quantified by the autocorrelation of each Fourier mode, $C_{\hat{u}}(k,t) = \langle \hat{u}(k,\tau)\hat{u}(k,\tau+t) \rangle$. 

For $\ell = 0~\unit{\um}$, the autocorrelation contours for both the nonlinear and linear cases exhibit qualitatively similar behavior, as shown in \cref{fig:2}(a): the lowest-wavenumber mode remains correlated up to $t \sim 10^{-3}~\unit{\ms}$, indicating slow relaxation, while modes with $k \gtrsim 30~\unit{\um}^{-1}$ decorrelate almost immediately. The contour plot visually highlights this trend through the steep narrowing of the correlated region as $k$ increases. This trend reflects the familiar $k^{2}$ dependence of viscous damping in the standard fluctuating hydrodynamics framework, where high-wavenumber modes relax much more quickly than large-scale modes. A more detailed comparison between the nonlinear and linear cases is shown for three representative wavenumbers: $k = 1$, $13$, and $126~\unit{\um}^{-1}$. At $k = 1~\unit{\um}^{-1}$, the autocorrelation remains close to unity up to $t \sim 10^{-4}~\unit{\ms}$, whereas at $k = 13~\unit{\um}^{-1}$ it has already dropped to roughly $0.2$ by this time. At the large wavenumber $k = 126~\unit{\um}^{-1}$, the autocorrelation is nearly zero throughout the entire simulation time span. This increasingly rapid decorrelation at higher wavenumbers is consistent with the trend observed in the contour plot. The corresponding autocorrelation curves for the three representative modes are nearly indistinguishable, demonstrating that the linear and nonlinear dynamics coincide when $\ell = 0~\unit{\um}$. This confirms that the linear approximation is valid in the uncorrelated case, as anticipated for standard fluctuating hydrodynamics.

For $\ell = 0.5~\unit{\um}$, the autocorrelation contours in \cref{fig:2}(b) show fundamentally different behavior between the nonlinear and linear cases, especially for $k \gtrsim 10~\unit{\um}^{-1}$ (roughly the wavenumber associated with the correlation length~$\ell$). The correlation time decreases for $k \lesssim 3~\unit{\um}^{-1}$, then increases up to $k \sim 10~\unit{\um}^{-1}$. Beyond this point, the correlation time decreases again in the nonlinear case, whereas in the linear case it increases beyond the simulation time span, indicating that the corresponding modes remain almost perfectly correlated throughout the simulation. This behavior directly reflects the exponential growth of $\Reyeff(k)$ in \cref{eq:Re_eff}, which predicts that nonlinear advection dominates viscous damping at sufficiently large wavenumbers when $\ell>0$.

The detailed comparison of the autocorrelation curves at the three representative wavenumbers reflects the same behavior. At $k = 1~\unit{\um}^{-1}$, the nonlinear and linear autocorrelation curves are nearly indistinguishable, and both are in close agreement with the corresponding 
curves in the uncorrelated case in \cref{fig:2}(a), suggesting that the linear approximation remains valid at small wavenumbers. At $k = 13~\unit{\um}^{-1}$, however, the linear autocorrelation remains close to unity throughout the simulation, while the nonlinear autocorrelation begins to decay around $t = 3 \times 10^{-3}~\unit{\ms}$. A similar trend holds at the large wavenumber $k = 126~\unit{\um}^{-1}$, where the nonlinear autocorrelation starts to decay around $t = 3 \times 10^{-4}~\unit{\ms}$, whereas the linear curve again remains near unity for the entire duration. This wavenumber-dependent divergence between the two cases is consistent with the prediction of \cref{sec:3} that, for finite correlation lengths, viscous damping is strongly suppressed at large $k$, leading to prolonged relaxation under the linear dynamics, while nonlinear mode coupling accelerates relaxation. Consequently, the effective Reynolds number $\Reyeff(k)$ grows large at high wavenumbers, rendering the linear approximation invalid at sufficiently small scales.

\subsection{Spatial correlations produce slower VACF decay and altered tracer diffusion\label{sec:4.2}}

\begin{figure}
    \centering
    \includegraphics[width=\linewidth]{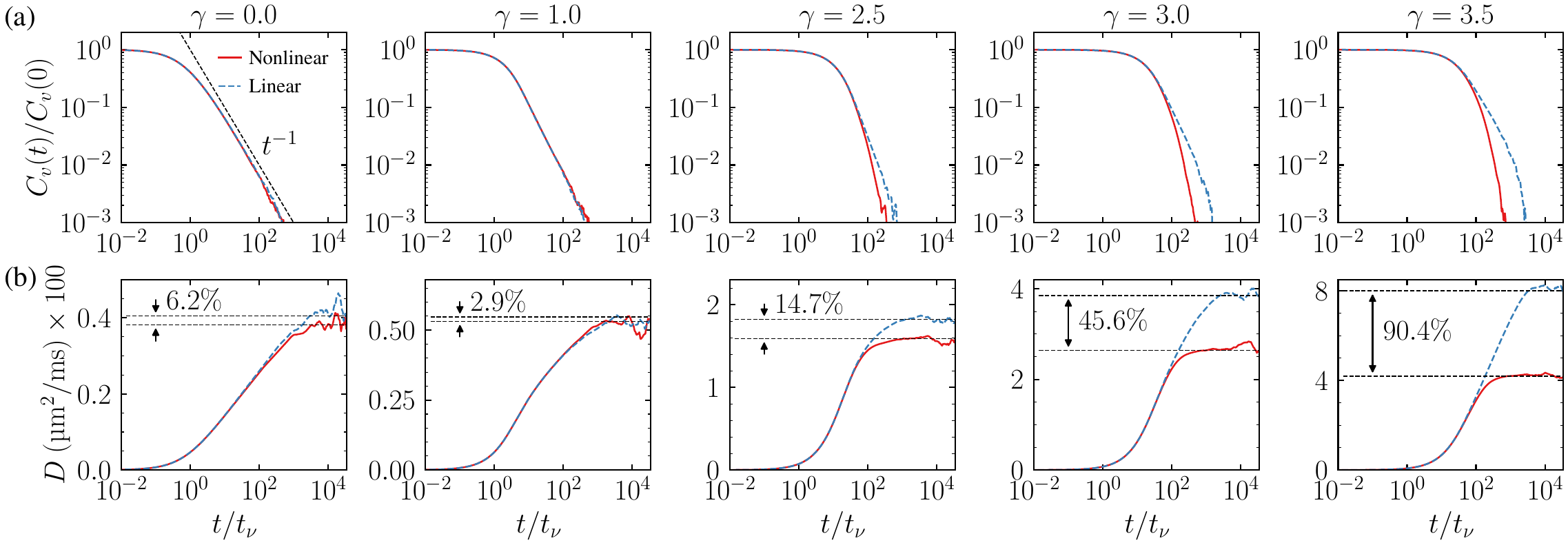}
    \caption{The particles exhibit stronger diffusion in the linear regime than in the nonlinear regime for increasing correlation length. This is manifested by (a) slower long-time decay of the VACFs and (b) larger diffusion coefficients. Results correspond to $\Rey \approx 0.4$, with viscous timescale $\tv = a^2/\nu$.}
    \label{fig:3}
\end{figure}

We next examine the two-dimensional case by analyzing tracer-particle diffusion under the correlated fluctuating Stokes and Navier--Stokes dynamics. This setting provides a more realistic hydrodynamic environment than the idealized one-dimensional Burgers system considered in the previous \cref{sec:4.1}. By comparing the VACFs produced by the linearized and nonlinear dynamics for different correlation lengths~$\ell$, we assess whether the behavior identified in one dimension persists in two dimensions and evaluate the validity of the linear Stokes approximation in this setting. For convenience, we define the dimensionless 
ratio $\gamma = \ell/a$.

\Cref{fig:3}(a) shows the particle VACF, $\autocorr(t) = \langle \bv(\tau)\bv(\tau+t)\rangle$, at five different correlation lengths~$\gamma$ and $\Rey\approx0.4$, where $\bv(t)$ denotes the particle velocity. The time axis is normalized by the viscous timescale $\tv = a^{2}/\nu$, which characterizes the time required for fluid momentum to diffuse across the particle radius~$a$~\cite{usabiaga2013jchemphys,usabiaga2014cmame}. For uncorrelated noise ($\gamma = 0$), the VACFs obtained from the nonlinear Navier--Stokes and linear Stokes equations agree closely. At short times ($t/\tv \lesssim 10^{-1}$), the VACF remains nearly flat and close to unity, corresponding to the ballistic regime. At intermediate times ($t/\tv \gtrsim 1$), the particle enters the hydrodynamic regime, where the VACF displays the classical $t^{-1}$ decay in two dimensions, reflecting the hydrodynamic long-time tail arising from momentum conservation~\cite{alder1970pra,ernst1970prl}. Thus, $\tv$ marks the timescale at which the VACF crosses over into this hydrodynamic regime. At long times ($t/\tv \gtrsim 10^{2}$), the VACF deviates from the $t^{-1}$ tail and decays exponentially due to the finite size of the periodic domain~\cite{atzberger2006,usabiaga2013jchemphys}. Overall, the recovery of the hydrodynamic long-time tail confirms that the simulations correctly capture the fluid--particle hydrodynamic coupling, and the close agreement between the nonlinear and linear cases shows that the linear Stokes approximation is valid for uncorrelated noise, again consistent with the predictions of standard fluctuating hydrodynamics.

For finite correlation lengths ($\gamma > 0$), the ballistic regime is sustained for significantly longer times than in the uncorrelated case ($\gamma = 0$), and its duration increases systematically with~$\gamma$. For example, at $\gamma = 1$ the ballistic regime persists up to $t/\tv \sim 1$, while for the largest value $\gamma = 3.5$ it extends to $t/\tv \sim 10$. This prolonged ballistic behavior reflects the fact that the spatial correlation in \cref{eq:lorentzian} hinders fluid momentum diffusion across scales: reduced viscous transport leads to more persistent velocity fluctuations in the underlying flow, which in turn extend the ballistic regime of tracer-particle motion~\cite{huang2025}. For $\gamma = 1$, the VACFs of the nonlinear and linear cases remain close over the entire time range. For larger $\gamma$, however, noticeable differences emerge: the VACF from the nonlinear dynamics decays more rapidly than its linear counterpart for $t/\tv \gtrsim 10^{2}$, with the discrepancy becoming increasingly pronounced as $\gamma$ increases. This breakdown is fully consistent with the prediction in \cref{sec:3}: for spatially correlated fluctuating hydrodynamics, viscous damping is strongly suppressed at high wavenumbers, causing the effective Reynolds number $\Reyeff(k)$ to grow large at small scales and rendering the scale-independent linear approximation invalid.

\begin{figure}
    \centering
    \includegraphics[width=\linewidth]{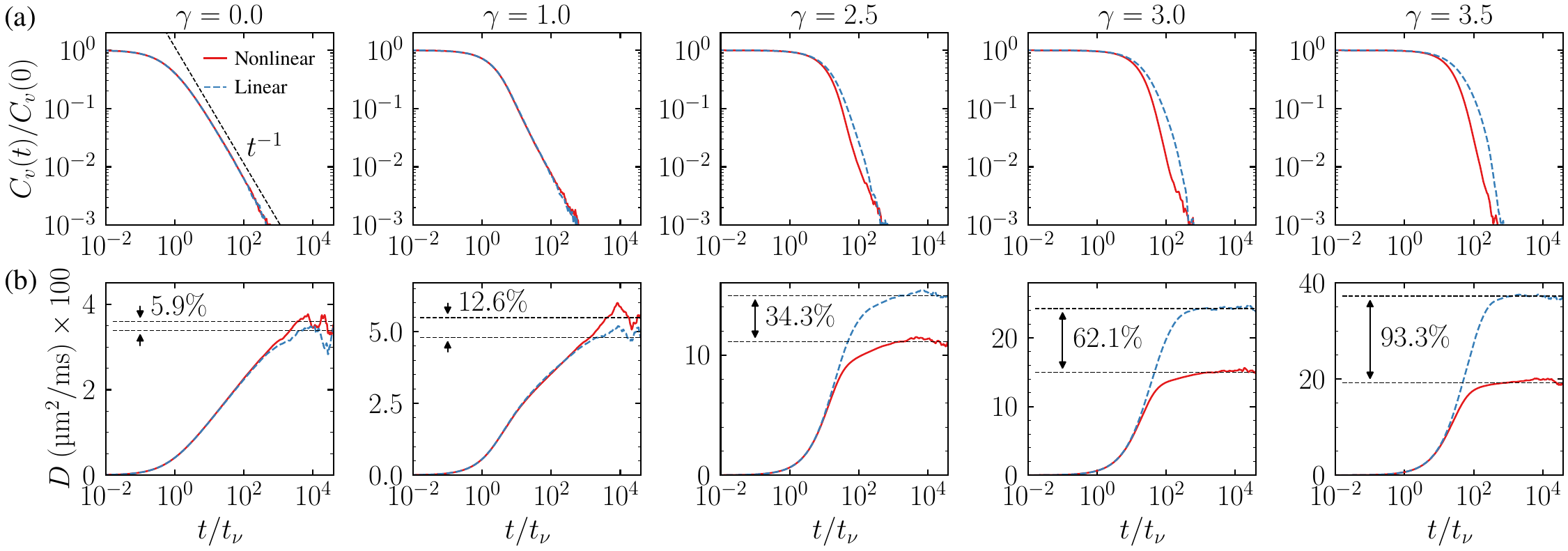}
    \caption{The particles again exhibit stronger diffusion in the linear regime than in the nonlinear regime for increasing correlation length, now at a higher Reynolds number $\Rey \approx 3$ than in \cref{fig:3}. The same qualitative behavior observed in \cref{fig:3} persists here, namely (a) a slower long-time decay of the VACFs and (b) larger diffusion coefficients with increasing correlation length. The viscous timescale is $\tv = a^2/\nu$.}
    \label{fig:4}
\end{figure}

To quantify the magnitude of the deviation between the nonlinear and linear cases, and thereby assess the severity of the breakdown of the linear approximation, we compare the time-dependent diffusion coefficient  \(D(t) = \int\autocorr(t')\d{t}'\), the time integral of the VACF. 
We measure the relative difference between the two cases using
\begin{equation}
    \frac{|\aab{D_{\text{linear}}(t)} - \aab{D_{\text{nonlinear}}(t)}|}{\aab{D_{\text{nonlinear}}(t)}} \times 100\%,
\end{equation}
where the average is taken over the time interval in which $D(t)$ has reached its plateau. As shown in \cref{fig:3}(b), the differences for $\gamma = 0$ and $\gamma = 1$ are approximately $6.2\%$ and $2.9\%$, respectively, both small enough to be attributed to statistical uncertainty. For larger correlation lengths, however, the discrepancies become substantial: the relative difference grows to $14.7\%$ at $\gamma = 2.5$, and reaches $90.4\%$ at $\gamma = 3.5$, where the linear prediction is nearly twice as large as the nonlinear result. These results confirm that the breakdown of the linear Stokes approximation becomes increasingly severe as the correlation length grows, consistent with the prediction in \cref{sec:3} that increasing correlation length further suppresses viscous damping at small scales.

At the higher Reynolds number $\Rey \approx 3$, the qualitative behavior observed at the lower $\Rey$ [\cref{fig:3}] persists, as shown in \cref{fig:4}. For uncorrelated noise ($\gamma = 0$), the VACFs obtained from the linear and nonlinear dynamics again agree closely and exhibit the expected hydrodynamic long-time decay. For finite correlation lengths, however, clear discrepancies emerge: the VACF from the nonlinear dynamics decays systematically faster than its linear counterpart at long times, with the separation becoming more pronounced as $\gamma$ increases. The time-dependent diffusion coefficient $D(t)$ exhibits the same trend: the relative difference between the linear and nonlinear plateau values grows with $\gamma$, reaching roughly $93.3\%$ at the largest correlation length considered ($\gamma = 3.5$). 

Taken together, these observations confirm a systematic failure of the linear approximation in the presence of finite spatial correlations and demonstrate that this breakdown persists across a range of conventional Reynolds numbers.


\section{Conclusions and Discussions\label{sec:5}}

We have demonstrated that the linear Stokes approximation, long regarded as valid whenever the conventional Reynolds number satisfies $\Rey \ll 1$, can fail when spatial correlations introduce scale-dependent viscous damping. In one dimension, we showed that for uncorrelated noise, the nonlinear and linear dynamics yield nearly identical relaxation of all Fourier modes, whereas for finite correlation length, the linear dynamics exhibit markedly slower relaxation of high-wavenumber modes than the full nonlinear dynamics. The two-dimensional particle-diffusion results lead to the same conclusion: at large correlation lengths, the diffusion coefficient predicted by the linear dynamics can differ from the nonlinear result by more than $90\%$. These findings are fully consistent with the predicted exponential growth of $\Reyeff(k)$ for correlated noise and confirm that nonlinear interactions remain essential even at small Reynolds numbers when a finite correlation length is present. More broadly, the loss of scale separation that accompanies nonlocal dissipation is a defining feature of correlated fluctuating hydrodynamics, and the framework developed here makes explicit how spatial correlations reshape the inertial--viscous balance across scales.

These observations also offer a useful perspective on a longstanding challenge in the study of complex and structured fluids: the difficulty of defining a meaningful, single-valued Reynolds number. In many such materials, the effective viscosity is not a fixed material parameter but depends on deformation rate, frequency, or even the relevant observation scale, reflecting the influence of microstructure, elasticity, or confinement. Although the present analysis focuses on equilibrium fluctuating fluids, the emergence of a scale-dependent Reynolds number suggests a possible route for generalizing inertial--viscous balance measures in systems where dissipation is redistributed across scales. The correlation length introduced here plays a role analogous to microstructural length scales found in polymer networks, gels, and dense suspensions, indicating how structural features might enter hydrodynamic descriptions through their impact on dissipation. In this sense, $\Reyeff(k)$ provides a conceptual prototype for formulating scale-dependent hydrodynamic control parameters in materials where the separation of scales underlying traditional Reynolds-number concepts is absent.

\section*{Acknowledgments}

The authors thank members of the Press\'e Lab for helpful discussions. This work was supported by the Army Research Office (ARO) under Grant No.~W911NF-23-1-0304. Numerical simulations were performed on the \textit{Sol} cluster at Arizona State University.


\bibliography{main}

@article{bandak2022,
    author  = {D. Bandak and N. Goldenfeld and A. A. Mailybaev and G. Eyink},
    journal = {Phys. Rev. E},
    title   = {Dissipation--range fluid turbulence and thermal noise},
    volume  = {105},
    pages   = {065113},
    year    = {2022},
    doi     = {10.1103/PhysRevE.105.065113}
}

@article{bandak2024,
    author  = {D. Bandak and A. A. Mailybaev and G. L. Eyink and N. Goldenfeld},
    journal = {Phys. Rev. Lett.},
    title   = {Spontaneous stochasticity amplifies even thermal noise to the largest scales of turbulence in a few eddy turnover times},
    volume  = {132},
    pages   = {104002},
    year    = {2024},
    doi     = {https://doi.org/10.1103/PhysRevLett.132.104002}
}

@book{landau1959,
    author    = {L. D. Landau and E. M. Lifshitz},
    publisher = {Pergamon Press},
    title     = {Fluid Mechanics},
    volume    = {6},
    series    = {Course of Theoretical Physics},
    year      = {1959},
    doi       = {https://doi.org/10.1016/C2013-0-03650-X}
}

@book{zarate2006,
    author    = {J. M. O. de Z\'arate and J. V. Sengers},
    publisher = {Elsevier},
    title     = {Hydrodynamic Fluctuations in Fluids and Fluid Mixtures},
    year      = {2006},
    doi       = {10.1016/B978-0-444-51515-5.X5000-5}
}

@book{kim1991,
    author    = {S. Kim and S. J. Karrila},
    publisher = {Butterworth-Heinemann},
    title     = {Microhydrodynamics: Principles and Selected Applications},
    year      = {1991},
    doi       = {https://doi.org/10.1016/C2013-0-04644-0}
}

@article{tamburrino2025,
    author  = {A. Tamburrino and Y. Ni\~no},
    journal = {Fluids},
    number  = {5},
    title   = {The universal presence of the {Reynolds} number},
    volume  = {10},
    pages   = {117},
    year    = {2025},
    doi     = {https://doi.org/10.3390/fluids10050117}
}

@article{berg1996,
    author  = {H. C. Berg},
    journal = {Proc. Natl. Acad. Sci. U. S. A.},
    number  = {25},
    title   = {Symmetries in bacterial motility},
    volume  = {93},
    pages   = {14225--14228},
    year    = {1996},
    doi     = {https://doi.org/10.1073/pnas.93.25.14225}
}

@article{groisman2004,
    author  = {A. Groisman and S. R. Quake},
    journal = {Phys. Rev. Lett.},
    number  = {9},
    title   = {A microfluidic rectifier: anisotropic flow resistance at low {Reynolds} numbers},
    volume  = {92},
    year    = {2004},
    pages   = {094501},
    doi     = {https://doi.org/10.1103/PhysRevLett.92.094501}
}

@article{qiu2014nc,
    author  = {T. Qiu and T.-C. Lee and A. G. Mark and K. I. Morozov and R. M\"unster and O. Mierka and S. Turek and A. M. Leshansky and P. Fischer},
    journal = {Nat. Commun.},
    title   = {Swimming by reciprocal motion at low {Reynolds} number},
    volume  = {5},
    pages   = {5119},
    year    = {2014},
    doi     = {https://doi.org/10.1038/ncomms6119}
}

@book{waigh2024cup,
    author    = {T. A. Waigh},
    publisher = {Cambridge University Press},
    title     = {The Physics of Bacteria: From Cells to Biofilms},
    year      = {2024},
    doi       = {https://doi.org/10.1017/9781009313506}
}

@article{salac2012jfm,
    author  = {D. Salac and M. J. Miksis},
    journal = {J. Fluid Mech.},
    number  = {25},
    title   = {Reynolds number effects on lipid vesicles},
    volume  = {711},
    pages   = {122--146},
    year    = {2012},
    doi     = {https://doi.org/10.1017/jfm.2012.380}
}

@article{davis1985arfm,
    author  = {R. H. Davis and A. Acrivos},
    journal = {Annu. Rev. Fluid Mech.},
    title   = {Sedimentation of noncolloidal particles at low {Reynolds} numbers},
    volume  = {17},
    pages   = {91--118},
    year    = {1985},
    doi     = {https://doi.org/10.1146/annurev.fl.17.010185.000515}
}

@article{he2024prf,
    author  = {N. He and Y. Cui and D. W. Q. Chin and T. Darnige and P. Claudin and B. Semin},
    journal = {Phys. Rev. Fluids},
    title   = {Sedimentation of a single soluble particle at low {Reynolds} and high {P\'eclet} numbers},
    volume  = {9},
    pages   = {044502},
    year    = {2024},
    doi     = {https://doi.org/10.1103/PhysRevFluids.9.044502}
}

@article{nguyen2005jfm,
    author  = {N.-Q. Nguyen and A. J. C. Ladd},
    journal = {J. Fluid Mech.},
    number  = {25},
    title   = {Sedimentation of hard-sphere suspensions at low {Reynolds} number},
    volume  = {525},
    pages   = {73--104},
    year    = {2005},
    doi     = {https://doi.org/10.1017/S0022112004002563}
}

@article{rallabandi2024arfm,
    author  = {B. Rallabandi},
    journal = {Annu. Rev. Fluid Mech.},
    title   = {Fluid--elastic interactions near contact at low {Reynolds} number},
    volume  = {56},
    pages   = {491--519},
    year    = {2024},
    doi     = {https://doi.org/10.1146/annurev-fluid-120720-024426}
}

@article{agarwal2025pof,
    author  = {S. Agarwal and A. C. Bekar and C. H\"uttig and D. S. Greenberg and N. Tosi},
    journal = {Phys. Fluids},
    title   = {Physics-based machine learning for mantle convection simulations},
    volume  = {37},
    pages   = {086624},
    year    = {2025},
    doi     = {https://doi.org/10.1063/5.0281832}
}

@article{brinkerhoff2021,
    author  = {D. Brinkerhoff and A. Aschwanden and M. Fahnestock},
    journal = {J. Glaciol.},
    title   = {Constraining subglacial processes from surface velocity observations using surrogate-basd {Bayesian} inference},
    volume  = {67},
    pages   = {385--403},
    year    = {2021},
    doi     = {https://doi.org/10.1017/jog.2020.112}
}

@article{pattyn2003,
    author  = {F. Pattyn},
    journal = {J. Geophys. Res. Solid Earth},
    title   = {A new three-dimensional higher-order thermomechanical ice sheet model: Basic sensitivity, ice stream development, and ice flow across subglacial lakes},
    volume  = {108},
    number  = {B8},
    pages   = {2382},
    year    = {2003},
    doi     = { https://doi.org/10.1029/2002JB002329}
}

@article{baskaran2009,
    author  = {A. Baskaran and M. C. Marchetti},
    journal = {Proc. Natl. Acad. Sci. U. S. A.},
    number  = {37},
    title   = {Statistical mechanics and hydrodynamics of bacterial suspensions},
    volume  = {106},
    pages   = {15567--15572},
    year    = {2009},
    doi     = {https://doi.org/10.1073/pnas.0906586106}
}

@article{brady1988,
    author  = {J. F. Brady and G. Bossis},
    journal = {Annu. Rev. Fluid Mech.},
    title   = {Stokesian dynamics},
    volume  = {20},
    pages   = {111--157},
    year    = {1988},
    doi     = {https://doi.org/10.1146/annurev.fl.20.010188.000551}
}

@article{phung1996,
    author  = {T. N. Phung and J. F. Brady and G. Bossis},
    journal = {J. Fluid Mech.},
    number  = {25},
    title   = {Stokesian dynamics simulation of {Brownian} suspensions},
    volume  = {313},
    pages   = {181--207},
    year    = {1996},
    doi     = {https://doi.org/10.1017/S0022112096002170}
}

@article{banchio2003,
    author  = {A. J. Banchio and J. F. Brady},
    journal = {J. Chem. Phys.},
    title   = {Accelerated {Stokesian dynamics: Brownian motion}},
    volume  = {118},
    pages   = {10323--10332},
    year    = {2003},
    doi     = {https://doi.org/10.1063/1.1571819}
}

@article{fiore2019,
    author  = {A. M. Fiore and J. W. Swan},
    journal = {J. Fluid Mech.},
    title   = {Fast {Stokesian} dynamics},
    volume  = {878},
    pages   = {544--597},
    year    = {2019},
    doi     = {https://doi.org/10.1017/jfm.2019.640}
}

@article{ouaknin2021,
    author  = {G. Y. Ouaknin and Y. Su and R. N. Zia},
    journal = {J. Comput. Phys.},
    title   = {Parallel accelerated {Stokesian} dynamics with {Brownian} motion},
    volume  = {442},
    pages   = {110447},
    year    = {2021},
    doi     = {https://doi.org/10.1016/j.jcp.2021.110447}
}

@article{michael2025pof,
    author  = {A. J. Michael and A. Mark and S. Sasic and H. Str\"om},
    journal = {Phys. Fluids},
    title   = {Generalized {Langevin} dynamics in multiphase direct numerical simulations using hydrodynamically optimized memory kernels},
    volume  = {37},
    pages   = {033317},
    year    = {2025},
    doi     = {https://doi.org/10.1063/5.0254930}
}

@article{hinch1975,
    author  = {E. J. Hinch},
    journal = {J. Fluid Mech.},
    title   = {Application of the {Langevin} equation to fluid suspensions},
    volume  = {72},
    pages   = {499--511},
    year    = {1975},
    doi     = {https://doi.org/10.1017/S0022112075003102}
}

@article{atzberger2007,
    author  = {P. J. Atzberger},
    journal = {Phys. D},
    number  = {2},
    title   = {A note on the correspondence of an immersed boundary method incorporating thermal fluctuations with {Stokesian--Brownian} dynamics},
    volume  = {226},
    pages   = {144--150},
    year    = {2007},
    doi     = {https://doi.org/10.1016/j.physd.2006.11.013}
}

@article{kavokine2021arfm,
    author  = {N. Kavokine and R. R. Netz and L. Bocquet},
    journal = {Annu. Rev. Fluid Mech.},
    title   = {Fluids at the nanoscale: from continuum to subcontinuum transport},
    volume  = {53},
    pages   = {377--410},
    year    = {2021},
    doi     = {https://doi.org/10.1146/annurev-fluid-071320-095958}
}

@article{usabiaga2013jchemphys,
    author  = {F. B. Usabiaga and X. Xie and R. Delgado-Buscalioni and A. Donev},
    journal = {J. Chem. Phys.},
    title   = {The {Stokes--Einstein} relation at moderate {Schmidt} number},
    volume  = {139},
    pages   = {214113},
    year    = {2013},
    doi     = {https://doi.org/10.1063/1.4834696}
}

@article{donev2014jsm,
    author  = {A. Donev and T. G. Fai and E. Vanden-Eijnden},
    journal = {J. Stat. Mech.},
    number  = {4},
    title   = {A reversible mesoscopic model of diffusion in liquids: from giant fluctuations to {Fick's} law},
    volume  = {2014},
    pages   = {P04004},
    year    = {2014},
    doi     = {https://doi.org/10.1088/1742-5468/2014/04/P04004}
}

@article{hauge1973jsp,
    author  = {E. H. Hauge and A. Martin-L\"of},
    journal = {J. Stat. Phys.},
    title   = {Fluctuating hydrodynamics and {B}rownian motion},
    volume  = {7},
    pages   = {259--281},
    year    = {1973},
    doi     = {https://doi.org/10.1007/BF01030307}
}

@article{delong2014jcp,
    author  = {S. Delong and F. B. Usabiaga and R. Delgado-Buscalioni and B. E. Griffith and A. Donev},
    journal = {J. Chem. Phys.},
    title   = {Brownian dynamics without {Green}'s functions},
    volume  = {140},
    pages   = {134110},
    year    = {2014},
    doi     = {https://doi.org/10.1063/1.4869866}
}

@article{noetinger1990,
    author  = {B. Noetinger},
    journal = {Physica A},
    title   = {Fluctuating hydrodynamics and {Brownian} motion},
    volume  = {163},
    pages   = {545--558},
    year    = {1990},
    doi     = {https://doi.org/10.1016/0378-4371(90)90144-H}
}

@article{delmotte2015,
    author  = {B. Delmotte and E. E. Keaveny},
    journal = {J. Chem. Phys.},
    title   = {Simulating {Brownian} suspensions with fluctuating hydrodynamics},
    volume  = {143},
    year    = {2015},
    pages   = {244109},
    doi     = {https://doi.org/10.1063/1.4938173}
}

@misc{huang2025,
    author       = {S. Huang and A. Saurabh and S. Press\'e},
    howpublished = {in revision},
    title        = {Spatially Correlated Noise Induces Transitions from the Diffusive to Ballistic Regime in Fluids},
    year         = {2025}
}

@article{usabiaga2013jcp,
    author  = {F. B. Usabiaga and L. Pagonabarraga and R. Delgado-Buscalioni},
    journal = {J. Comput. Phys.},
    title   = {Inertial coupling for point particle fluctuating hydrodynamics},
    volume  = {235},
    pages   = {701--722},
    year    = {2013},
    doi     = {10.1016/j.jcp.2012.10.045}
}

@article{usabiaga2014cmame,
    author  = {F. B. Usabiaga and R. Delgado-Buscalioni and B. E. Griffith and A. Donev},
    journal = {Comput. Methods Appl. Mech. Engrg.},
    title   = {Inertial coupling method for particles in an incompressible fluctuating fluid},
    volume  = {269},
    pages   = {139--172},
    year    = {2014},
    doi     = {10.1016/j.cma.2013.10.029}
}

@article{delong2013pre,
    author  = {S. Delong and B. E. Griffith and E. Vanden-Eijnden and A. Donev},
    journal = {Phys. Rev. E},
    title   = {Temporal integrators for fluctuating hydrodynamics},
    volume  = {87},
    pages   = {033302},
    year    = {2013},
    doi     = {10.1103/PhysRevE.87.033302}
}

@article{donev2010camcs,
    author  = {A. Donev and E. Vanden-Eijnden and A. Garcia and J. Bell},
    journal = {Comm. App. Math. and Comp. Sci.},
    number  = {2},
    title   = {On the accuracy of finite-volume schemes for fluctuating hydrodynamics},
    volume  = {5},
    pages   = {149--197},
    year    = {2010},
    doi     = {https://doi.org/10.2140/camcos.2010.5.149}
}

@misc{supplemental,
    author       = {},
    howpublished = {},
    title        = {See {Supplemental Material} at [URL will be inserted by publisher] for simulation details, which includes {Ref}.~[].},
    year         = {}
}

@article{usabiaga2012mms,
    author  = {F. B. Usabiaga and J. B. Bell and R. Delgado-Buscalioni and A. Donev and T. G. Fai and B. E. Griffith and C. S. Peskin},
    journal = {Multiscale Model. Simul.},
    issue   = {4},
    title   = {Staggered schemes for fluctuating hydrodynamics},
    volume  = {10},
    pages   = {1369--1408},
    year    = {2012},
    doi     = {10.1137/120864520}
}

@book{canuto2006,
    author    = {C. Canuto and M. Y. Hussaini and A. Quarteroni and T. A. Zang},
    publisher = {Springer},
    title     = {Spectral Methods: Fundamentals in Single Domains},
    year      = {2006},
    doi       = {https://doi.org/10.1007/978-3-540-30726-6}
}

@book{canuto2007,
    author    = {C. Canuto and M. Y. Hussaini and A. Quarteroni and T. A. Zang},
    publisher = {Springer},
    title     = {Spectral Methods: Evolution to Complex Geometries and Applications to Fluid Dynamics},
    year      = {2007},
    doi       = {https://doi.org/10.1007/978-3-540-30728-0}
}

@book{oksendal2003,
    author    = {B. {\O}ksendal},
    publisher = {Springer},
    title     = {Stochastic Differential Equations: An Introduction with Applications},
    edition   = {6},
    year      = {2003},
    doi       = {https://doi.org/10.1007/978-3-642-14394-6}
}

@article{alder1970pra,
    author  = {B. J. Alder and T. E. Wainwright},
    journal = {Phys. Rev. A},
    title   = {Decay of the velocity autocorrelation function},
    volume  = {1},
    pages   = {18--21},
    year    = {1970},
    doi     = {10.1103/PhysRevA.1.18}
}

@article{ernst1970prl,
    author  = {M. H. Ernst and E. H. Hauge and J. M. J. van Leeuwen},
    journal = {Phys. Rev. Lett.},
    number  = {18},
    title   = {Asymptotic time behavior of correlation functions},
    volume  = {25},
    pages   = {1254--1256},
    year    = {1970},
    doi     = {10.1103/PhysRevLett.25.1254}
}

@article{atzberger2006,
    author  = {P. J. Atzberger},
    journal = {Phys. Lett. A},
    number  = {4--5},
    title   = {Velocity correlations of a thermally fluctuating {Brownian} particle: a novel model of the hydrodynamic coupling},
    volume  = {351},
    pages   = {225-230},
    year    = {2006},
    doi     = {https://doi.org/10.1016/j.physleta.2005.10.107}
}

@inproceedings{cufinufft,
    author    = {Y.-H. Shih and G. Wright and J. And\'en and J. Blaschke and A. H. Barnett},
    booktitle = {PDSEC2021 workshop of the IPDPS2021 conference},
    title     = {{cuFINUFFT: a load-balanced GPU library for general-purpose nonuniform FFTs}},
    year      = {2021},
    doi       = {10.1109/IPDPSW52791.2021.00105}
}

@article{saldana2024fluids,
    author  = {M. Saldana and S. Gallegos and E. G\'alvez and J. Castillo and E. Salinas-Rodriguez and E. Cerecedo-S\'aenz and J. Hern\'andez-\'Avila and A. Navarra and N. Toro},
    journal = {Fluids},
    number  = {12},
    title   = {The {Reynolds} number: a journey from its origin to modern Applications},
    volume  = {9},
    pages   = {299},
    year    = {2024},
    doi     = {https://doi.org/10.3390/fluids9120299}
}

@article{grimm2025sm,
    author    = {N. Grimm and J. Baschnagel and A. N. Semenov and A. Zippelius and M. Fuchs},
    journal   = {Soft Matter},
    publisher = {Royal Society of Chemistry (RSC)},
    title     = {Stress correlations and stress memory kernels in viscoelastic fluids},
    volume    = {21},
    pages     = {4256--4274},
    year      = {2025},
    doi       = {10.1039/D5SM00156K}
}

@book{batchelor2000,
    author    = {G. K. Batchelor},
    publisher = {Cambridge University Press},
    title     = {An Introduction to Fluid Dynamics},
    year      = {2000},
    doi       = {https://doi.org/10.1017/CBO9780511800955}
}

@article{turk2024jfm,
    author  = {G. Turk and R. Adhikari and R. Singh},
    journal = {J. Fluid Mech.},
    title   = {Fluctuating hydrodynamics of an autophoretic particle near a permeable interface},
    volume  = {998},
    pages   = {A34},
    year    = {2024},
    doi     = {https://doi.org/10.1017/jfm.2024.661}
}

@article{delmotte2025,
    author  = {B. Delmotte and F. B. Usabiaga},
    journal = {Phys. Rev. Fluids},
    title   = {Modeling complex particle suspensions: perspectives on the rigid multiblob method},
    volume  = {10},
    pages   = {100701},
    year    = {2025},
    doi     = {https://doi.org/10.1103/64b1-lfmc}
}

@article{seyler2019,
    author  = {S. L. Seyler and S. Press\'e},
    journal = {Phys. Rev. Res.},
    title   = {Long-time persistence of hydrodynamic memory boosts microparticle transport},
    volume  = {1},
    pages   = {032003(R)},
    year    = {2019},
    doi     = {https://doi.org/10.1103/PhysRevResearch.1.032003}
}

@article{seyler2020,
    author  = {S. L. Seyler and S. Press\'e},
    journal = {J. Chem. Phys.},
    title   = {Surmounting potential barriers: hydrodynamic memory hedges against thermal fluctuations in particle transport},
    volume  = {153},
    pages   = {041102},
    year    = {2020},
    doi     = {https://doi.org/10.1063/5.0013722}
}

\end{document}


\begin{CJK*}{UTF8}{gbsn}

\title{
Supplemental Material: Revisiting the scale-dependence of the Reynolds number in correlated fluctuating fluids 
}

\author{Sijie Huang (黄斯杰)}
\author{Ayush Saurabh}
\affiliation{
Department of Physics, Arizona State University, Tempe, AZ 85287, USA
}
\affiliation{
Center for Biological Physics, Arizona State University, Tempe, AZ 85287, USA
}
\author{Steve Press\'e}
\email{Corresponding author: spresse@asu.edu}
\affiliation{
Department of Physics, Arizona State University, Tempe, AZ 85287, USA
}
\affiliation{
Center for Biological Physics, Arizona State University, Tempe, AZ 85287, USA
}
\affiliation{
School of Molecular Sciences, Arizona State University, Tempe, AZ 85287, USA
}
\date{\today}
\maketitle

\end{CJK*}

\section{Simulation details}

In this section, we provide the full information on the numerical simulations discussed in the main text. We consider spatially correlated fluctuating hydrodynamics in both one and two dimensions. In two dimensions, we simulate the spatially correlated fluctuating incompressible Navier--Stokes equation that governs the fluid phase
\begin{equation}
    \label{eq:corr_fluc_ns}
    \partial_t\bu + \bu\cdot\grad\bu = -\grad p + \divergence(\nueff*\grad\bu) + \sqrt{2\nu\kBT\rho^{-1}}\,\divergence\bcZ,
\end{equation}
where $\bcZ$ is the spatially correlated noise defined in the main text, \(\nueff(r) = \nu\corr(r)\) is the effective viscosity, and \(\corr(r)\) is the spatial correlation function that depends only on the two-point separation \(r=|\bx-\bx'|\). In one dimension, we consider the following one-dimensional fluctuating Burgers equation
\begin{equation}
    \label{eq:corr_fluc_burgers}
    \partial_tu + u\partial_xu = \partial_x(\nueff*\partial_xu) + \sqrt{2\nu\kBT\rho^{-1}}\,\partial_x\eta,
\end{equation}
where the noise covariance is given by
\begin{equation}
    \aab*{\eta(x,t)\eta(x',t')} = \corr(r)\delta(t-t').
\end{equation}
The linear counterparts of \cref{eq:corr_fluc_ns,eq:corr_fluc_burgers} can be obtained by neglecting the nonlinear terms. Recall that we consider the following Lorentzian correlation function in the main text 
\begin{equation}
    \corr(r) \propto \frac{\ell}{(\ell^2 + r^2)^{(d+1)/2}},\quad \corrhat(k)\propto e^{-\ell k},
\end{equation}
where $d$ is the spatial dimension, and $\corrhat(k)$ is its Fourier transform. 

In the following, we first describe the spatiotemporal discretization strategy employed in this work in \cref{sec:1.1}, using the one-dimensional stochastic Burgers equation [\cref{eq:corr_fluc_burgers}] as an illustrative example. We then present, in \cref{sec:1.2}, the fluid--particle coupling method used to simulate particle diffusion in two-dimensional correlated fluctuating fluids. The same spatiotemporal discretization strategy adopted in the one-dimensional case is applied to the two-dimensional spatially correlated Navier--Stokes equation [\cref{eq:corr_fluc_ns}]; however, the fluid--particle coupling requires additional discussion and is therefore described in detail.

\subsection{Spatiotemporal discretization of the stochastic Burgers equation\label{sec:1.1}}

In this work, the one-dimensional stochastic Burgers equation in \cref{eq:corr_fluc_burgers} is solved numerically on a periodic domain. Periodic boundaries mimic an unbounded, homogeneous fluid, eliminating boundary fluxes and ensuring that no energy is injected or removed. They therefore provide the simplest boundary conditions consistent with the FDR, and require no additional boundary corrections~\cite{usabiaga2012mms,delong2014jcp}. \Cref{eq:corr_fluc_burgers} is spatially discretized using a Fourier--Galerkin pseudospectral method with $3/2$-rule de-aliasing to eliminate the aliasing errors arising from the nonlinear term~\cite{canuto2007}. This pseudospectral discretization provides spectral accuracy and conserves energy at the discrete level, thereby preventing spurious energy injection or dissipation associated with violations of the fluctuation--dissipation relation (FDR).

Special care is required in the spatiotemporal discretization of the stochastic forcing. We interpret the discretized correlated noise following the same interpretation of discretized white noise introduced in Ref.~\cite{donev2010camcs}. Specifically, the discrete noise is understood as a spatiotemporal average of the underlying continuous noise over each computational cell and time step, which introduces a normalization factor $(\Delta V\Delta t)^{-1/2}$, where $\Delta V$ and $\Delta t$ denote the cell volume and time step size, respectively. Accordingly, the discretized correlated noise in Fourier space is defined as
\begin{equation}
    \hat{\eta}(k) = \frac{\corrhat(k)}{(\Delta V\Delta t)^{1/2}}\hat{\xi}(k),
\end{equation}
where $\hat{\xi}$ denotes a set of independent standard normal random variables. The factor $(\Delta V\Delta t)^{-1/2}\hat{\xi}$ therefore represents discretized spatially white noise, and multiplication by the correlation function $\corrhat(k)$ yields the desired correlated noise via the convolution theorem.

For temporal integration, the nonlinear term is advanced in time using Heun's method, which has been shown to perform robustly in fluctuating hydrodynamics simulations~\cite{usabiaga2012mms,delong2013pre}. The linear viscous term and the stochastic forcing are treated exactly via integrating factors~\cite{canuto2006,oksendal2003}. The resulting time-stepping scheme reads
\begin{subequations}
    \begin{align}
        \uhat^{\star,n+1} &= \Phi(\uhat^n + \Delta tN^n) + \iu\sqrt{\Theta}\Gamma k\hat{\eta}^n, \\
        \uhat^{n+1} &= \Phi\uhat^n + \frac{\Delta t}{2}(\Phi N^n + N^{\star,n+1}) + \iu\sqrt{\Theta}\Gamma k\hat{\eta}^n,
    \end{align}
\end{subequations}
where the two integrating factors $\Phi$ and $\Gamma$ are given by 
\begin{equation}
    \Phi(k) = e^{\Delta tL(k)},\quad \Gamma(k) = \sqrt{\frac{e^{2\Delta tL(k)} - 1}{2L(k)}},
\end{equation}
where $L(k) = -\nu k^2\corrhat(k)$ is the Fourier transform of the viscous linear operator.

\subsection{Fluid--particle coupling in two dimensions\label{sec:1.2}}

In two dimensions, we consider tracer particles diffusing in the fluctuating correlated fluid in a doubly periodic domain of side length $4\pi~\unit{\um}$. This is inherently a fluid--structure interaction problem. To couple the fluid and particle phases, we follow the general strategy of an immersed boundary method developed in Refs.~\cite{usabiaga2013jcp,usabiaga2014cmame}. The method imposes a no-slip condition on the particle surface, enforcing zero relative velocity between the fluid and the particle, thereby conserving the total momentum of the combined system. When the particles are treated as passive and inertialess, they are advected passively by the flow without exerting forces back on the fluid. Their motion reduces to 
\begin{subequations}
    \label{eq:sim_particle}
    \begin{align}
        \odv*{\bX(t)}{t} &= \bv(\bX(t)), \label{eq:sim_particle_motion} \\
        \bv(\bX(t)) &= \bJ(\bX(t))\bu(\bx,t) = \int\delta_a(\bX(t) - \bx)\bu(\bx,t)\d{\bx}, \label{eq:sim_interpolation}
    \end{align}
\end{subequations}
where $\bu(\bx,t)$ is the flow velocity field governed by \cref{eq:corr_fluc_ns}, $\bX(t)$ is the particle position, $\bv(\bX(t))$ its velocity, and $\delta_a$ an averaging kernel of width $a$ that also defines the particle's volume. The operator $\bJ(\bX(t))$ averages $\bu(\bx)$ around \(\bX(t)\), so that $\bv(\bX(t))$ represents the locally averaged fluid velocity experienced by the particle. Therefore, \cref{eq:sim_particle} describes a passive tracer advected by the coarse-grained flow field \(\bv\)~\cite{donev2014jsm}.

The fluid momentum equation \cref{eq:corr_fluc_ns} is solved using the same discretization as the one-dimensional Burgers equation described in \cref{sec:1.1}. Particle trajectories described by \cref{eq:sim_particle} are integrated in time using a midpoint predictor--corrector scheme~\cite{usabiaga2013jcp,usabiaga2014cmame}. The resulting discretized equations are given by 
\begin{subequations}
    \label{eq:discretized_system}
    \begin{align}
        \bX^{\star,n+1} &= \bX^n +\frac{\Delta t}{2}\bJ^n\bu^n, \\
        \buhat^{\star,n+1} &= \Phi\pab{\buhat^n + \Delta t\bcP\bN^{n}} + \iu\sqrt{\Theta}\Gamma\bcP\bk\cdot\widehat{\bm{Z}}^n, \\
        \buhat^{n+1} &= \Phi\buhat^n + \frac{\Delta t}{2}\bcP\pab{\Phi\bN^{n} + \bN^{\star,n+1}} + \iu\sqrt{\Theta}\Gamma\bcP\bk\cdot\widehat{\bm{Z}}^n, \\
        \bX^{n+1} &= \bX^{n} + \Delta t\bJ^{\star,n+1}\pab{\frac{\bu^{n+1} + \bu^n}{2}},
    \end{align}
\end{subequations}
where \(\Delta t\) is the timestep size, 
\[
\Phi(\bk) = e^{\Delta tL(k)},
\quad 
\Gamma(\bk)=\sqrt{\frac{e^{2\Delta tL(k)} - 1}{2L(k)}},
\]
$\bN^n = -\widehat{\bu^n\grad\bu^n}$, $\bJ^n = \bJ(\bX^n)$, and $\Theta=2\nu\kBT/(\rho\Delta V_\mathrm{f}\Delta t)$. The random stress tensor \(\widehat{\bm{Z}}(\bk)\) is the discrete correlated noise following the spatiotemporal discretization described in \cref{sec:1.1}. To generate this correlated noise efficiently, spatially white noise is multiplied by the prescribed spectral correlation function,
\begin{equation}
    \widehat{\bm{Z}}(\bk) = \corrhat(\bk)\frac{\widehat{\bm{W}} + \widehat{\bm{W}}^\top}{\sqrt{2}},
\end{equation}
which avoids costly real-space convolutions and, through symmetrization, preserves angular momentum~\cite{usabiaga2012mms}. 

The particle tracking strategy follows Ref.~\cite{donev2014jsm}. The operation $\bJ(\bX(t))$ in \cref{eq:sim_particle} consists of two steps: the fluid velocity field is first low-pass filtered in Fourier space, and then the particle velocity is obtained by interpolating the filtered field at the particle position $\bX(t)$. In the present work, the following Gaussian function is used as the filter kernel to approximate Peskin's four-point kernel~\citep{usabiaga2013jcp}
\begin{equation}
    G_a(\bx) = (\pi a^2)^{-d/2}\exp\pab{-\frac{\bx^2}{a^2}},
\end{equation}
where $a$ is assumed to be the particle radius. The volume of the particle $\Delta V_\mathrm{p}$ can be approximated by $\Delta V_{\mathrm{p}} = \bab{\int G_a^2(\bx)\d{\bx}}^{-1} = (2\pi a^2)^{d/2}$~\cite{usabiaga2013jcp,usabiaga2014cmame}. After filtering, the interpolation is carried out using a GPU-based non-uniform fast Fourier transform~\cite{cufinufft}.

\subsection{Simulation conditions}

In this section, we summarize the simulation setup and parameter choices. The one-dimensional system defined in \cref{eq:corr_fluc_burgers} is solved on a periodic domain of length $4\pi~\unit{\um}$, discretized using $512$ uniformly spaced grid points. The kinematic viscosity is set to $\nu=875~\unit{\um^2/\ms}$, corresponding to that of water at a temperature of \qty{300}{\K}. 

The two-dimensional fluid--particle system described by \cref{eq:corr_fluc_ns,eq:sim_particle} is solved on a doubly periodic domain with side length $4\pi~\unit{\um}$, discretized on a $1024^2$ uniform grid. In addition to the water-like viscosity, we also consider a lower viscosity value, $\nu = 100~\unit{\um^{2}/\ms}$, representative of liquid metals such as mercury at approximately \qty{293}{\K}. The particle radius is set to $a=\qty{0.05}{\um}$, approximately four times the grid spacing.

\begin{table}
    \centering
    \renewcommand{\arraystretch}{1.5}
    \begin{tabular}{c|ccccc}
        \hline
        $d$& $\nu~(\unit{\um^2/\ms})$ & $L~(\unit{\um})$  & $\ell~(\unit{\um})$ & $\Delta x~(\unit{\um})$ & $\Delta t~(\unit{\ms})$  \\
        \hline
        $1$ & $875$ & $4\pi$ & $0,0.5$ & $0.025$ & $10^{-5}$ \\
        \hline 
        \multirow{2}{*}{$2$} & $875$ & $4\pi$ & $0,0.05,0.125,0.15,0.175$ & $0.012$ & $5\times10^{-8}$  \\
        & $100$ & $4\pi$ & $0,0.05,0.125,0.15,0.175$ & $0.012$ & $5\times10^{-7}$ \\ 
        \hline 
    \end{tabular}
    \caption{Simulation parameters. $d$ is the spatial dimensions, $\nu$ is the kinematic viscosity, $L$ is the domain size, $\ell$ is the correlation length, $\Delta x$ and $\Delta t$ are the grid spacing and timestep size, respectively.}
    \label{tab:sim_params}
\end{table}

The simulation parameters used in this work are provided in \cref{tab:sim_params}. All simulations were performed on GPUs using \texttt{CuPy}\footnote{v.13.6.0, \url{https://cupy.dev/}.}.


\bibliography{main}